\documentclass[ALICE,manyauthors]{cernphprep}

\usepackage[comma,square,numbers,sort&compress]{natbib}
\usepackage{hyperref}
\usepackage{lineno}
\usepackage{color}
\begin{document}%

%%%%%%%%%%%%%%%  Title page %%%%%%%%%%%%%%%%%%%%%%%%
\begin{titlepage}
%
%\PHnumber{2017-266}    % required, will be obtained from PH
\PHdate{4 October}              % required, will be obtained from PH
\PHyear{2017}
\PHnumber{266}      % required, will be obtained from PH
%\PHdate{\monthname\,\the\day,}  % required, will be obtained from PH

%%% Put your own title + short title here:
\title{Production of $^{4}$He and $^{4}\overline{\textrm{He}}$ in Pb--Pb collisions at $\sqrt{s_{\mathrm{NN}}}$ = 2.76 TeV \\ at the LHC}
\ShortTitle{Production of $^{4}$He and $^{4}\overline{\textrm{He}}$}   % appears on right page headers

%%% Do not change the next lines
\Collaboration{ALICE Collaboration\thanks{See Appendix~\ref{app:collab} for the list of collaboration members}}
\ShortAuthor{ALICE Collaboration} % appears on left page headers, do not change

\begin{abstract}
Results on the production of $^{4}$He and $^{4}\overline{\textrm{He}}$ nuclei in Pb--Pb collisions at $ \sqrt{s_{\mathrm{NN}}} = 2.76 $~TeV in the rapidity range $ \mid y \mid < 1$, using the ALICE detector, are presented in this paper. The rapidity densities corresponding to 0-10\% central events are found to be \mbox{$\mathrm{d}N/\mathrm{d}y _{^{4}\mathrm{He}} =  (0.8  \pm 0.4 ~(\mathrm{stat}) \pm 0.3~(\mathrm{syst}))\times 10^{-6}$} and \mbox{$\mathrm{d}N/\mathrm{d}y _{^{4}\mathrm{\overline{He}}} =  (1.1  \pm 0.4~(\mathrm{stat}) \pm 0.2~(\mathrm{syst}))\times 10^{-6}$}, respectively. This is in agreement with the statistical thermal model expectation assuming the same chemical freeze-out temperature\\ \mbox{($T_{\mathrm{chem}}$ = 156~MeV)} as for light hadrons. The measured ratio of $^{4}\overline{\mathrm{He}}$/$^{4}$He is $1.4 \pm 0.8~(\mathrm{stat}) \pm 0.5~(\mathrm{syst})$.
\end{abstract}
\end{titlepage}
\setcounter{page}{2}

\section{Introduction}
\label{sec:Intro}

The production of light (hyper-)nuclei, up to a mass number $A=3$, has been reported already in Pb--Pb collisions at $\sqrt{s_{\mathrm{NN}}}$ = 2.76 TeV at the Large Hadron Collider (LHC). This includes deuterons, $^{3}$He and the hypertriton as well as their corresponding anti-particles~\cite{nuclei,hypertriton}.  %\textcolor{red}{Was ist mit anti-tritons? Weil die haben wir ja observed aber keinen yield}. 
The observed total yields can be described well by equilibrium thermal models~\cite{pbm1,pbm,thermalModel,thermalModel1,qm2012_pbm,anton1,steinheimer}, with only three free parameters: the chemical freeze-out temperature $T_{\mathrm{chem}}$, the volume $V$ and the baryo-chemical potential $ \mu _{B}$. The current best fit to the measured yields at the LHC, including results ranging in mass from pions up to $ ^{3} $He, results in a $T_{\mathrm{chem}} = 156$ MeV~\cite{QM2014Michele}. The measurement of the production yields of $^4$He and $^{4}\overline{\mathrm{He}}$ ($A=4$) will put additional constraints on $T_{\mathrm{chem}}$. 
Since the baryo-chemical potential is consistent with zero ($\mu _{B} = 0.7 \pm 3.8$ MeV \cite{Andronic:2016nof}) at LHC energies, the expected anti-baryon to baryon ratio is unity. Therefore, also the ratio is expected to be close to unity for particles composed of (anti-)baryons, namely the anti-nuclei and nuclei~\cite{thermalModel1}.  

Furthermore, $^{4}\overline{\mathrm{He}}$ is the heaviest anti-nucleus ever observed. It was discovered in Au--Au collisions at RHIC by the STAR collaboration~\cite{star_alpha}. Out of 10$ ^{9} $ Au--Au collisions at centre-of-mass energies per nucleon pair ($\sqrt{s_{\mathrm{NN}}}$) of 200 GeV and 62.4 GeV, 18 $ ^{4}\overline{\mathrm{He}} $ have been detected. The corresponding yield at a given transverse momentum $p_{\mathrm{T}}$ is compared to the prediction of the thermal model~\cite{pbm2007} and the coalescence nucleosynthesis model~\cite{Sato} and found to be consistent with both. %Werte im thermal model können nicht mit dN/dy bei bestimmtem pT verglichen werden
A confirmation of this observation is still pending as no other experiment has been able to detect the $^{4}\overline{\mathrm{He}}$ particle since then.

Coalescence models have been successfully used to describe the general trends of deuteron production~\cite{Hagedorn, butler_pearson61, butler_pearson63, Nagle, Scheibl, Ko1, Ko2, Ko3, Ko4, Botvina, Sun} in relativistic nuclear collisions, albeit with a number of external parameters. These models are clearly challenged with the regular pattern observed in the production probability for light nuclei measured by the STAR~\cite{star_alpha} and ALICE~\cite{nuclei} Collaborations. To extend the studies to A=4 the measurement at LHC energies is obviously of great interest.

In this paper, the measurement of the production yield of the $^4$He and $^{4}\overline{\mathrm{He}}$ nuclei with the ALICE apparatus is presented. Besides the increase in collision energy, the main difference with respect to the measurement by the STAR Collaboration is the usage of a six layer silicon vertex detector in ALICE. Together with the other barrel detectors this provides precision information on vertex position, particle identification and momentum. The determined yields are compared to thermal model expectations.

\section{Detector setup and Data sample}
\label{sec:detector}

The two main detectors involved in the identification of the $^4$He and $^{4}\overline{\mathrm{He}}$ particles are the Time Projection Chamber (TPC)~\cite{alice_tpc} and the Time of Flight (TOF) detector~\cite{alice_tof}, combined with the start time detector T0. In addition, V0 detectors (\cite{vzero2,alice_centrality}) are used for centrality determination and the Inner Tracking System (ITS)~\cite{alice_its} is employed for tracking and the discrimination between primary and secondary particles~\cite{nuclei,PrimaryParticles}. A full description of the ALICE detector can be found in~\cite{alice}, whereas the performance of the ALICE sub-detectors is reported in~\cite{alice_performance}. 

The measurement of the $^4$He and $^{4}\overline{\mathrm{He}}$ particles is performed on the 2011 data set of Pb--Pb collisions at  $ \sqrt{s_{\mathrm{NN}}} = 2.76 $ TeV. From this campaign, $38.7 \times 10^{6} $ events in a trigger mix of central, semi-central and minimum-bias events are used in this analysis. This leads to $20.7 \times 10^{6} $ events in the 0-10\% centrality interval,  $17.4 \times 10^{6} $ events in the 10-50\% centrality interval and $0.6 \times 10^{6} $ events in the 50-80\% centrality interval. The combined yields are extrapolated to the 0-10\% centrality class with the procedure discussed in section 4.

\section{Analysis}

To ensure high tracking efficiency, high energy-deposit $\left( \mathrm{d}E/\mathrm{d}x \right) $ resolution in the TPC and a good track matching between the TPC and TOF detectors, a set of selection criteria is applied. %Each track is required to have at least 80 of up to 159 clusters in the TPC attached to it, with a  $\chi^2$ of the momentum fit that is smaller than 4 per cluster. To achieve final precision the accepted tracks are refit while the track finding algorithm is run inwards, outwards and back (for more details on the ALICE tracking see \cite{alice_performance} and section 5 of~\cite{PPR}). In the ITS at least 2 of up to 6 hits are required. 
In order to select primary particles, the corresponding tracks have to originate from the primary vertex. The primary vertex position is estimated using the ITS and the TPC detectors. The resolution of the vertex determination is better than 50~$\mu$m in the xy-plane and 150 $\mu$m in the z-direction for charged particles with momenta above 1 GeV/$c$.
To select primary tracks, the minimum distance from the vertex, called Distance-of-Closest-Approach (DCA), is required to be smaller than 1 cm along the z-axis, whereas the DCA in the xy-plane must not be greater than 0.1 cm. In addition,  a hit in the TOF detector is required for a precise time measurement and only those tracks are used for the track reconstruction. 
The selection criteria are summarised in Table~\ref{tab:cuts}.

\begin{table}
  \centering
  \begin{tabular}{l | c }
    %{\bf cut} & {\bf value} \\
    \hline
    {\bf Track selection criteria} & {\bf value}  \\
    \hline
    Number of clusters in TPC & $n_{\mathrm{cl}} > 80$ \\
    Number of hits in ITS & $n_{\mathrm{hits}} > 2$ \\  
    TPC track quality & $\chi^{2}/\mathrm{cluster} < 4$ \\
    Acceptance in pseudo-rapidity   & $ \vert \eta \vert < 0.8 $ \\
    Acceptance in rapidity & $ \vert y \vert < 1 $\\
    DCA$ _{z} $&  DCA$ _{z} $ $< 1 \; \mathrm{cm}$  \\ 
    DCA$ _{xy} $&  DCA$ _{xy} $ $< 0.1 \; \mathrm{cm}$  \\ 
    \hline 
    {\bf PID selection} &  {\bf value}\\
    \hline
    TPC PID cut &  $\pm$3$\sigma $ \\ 
    TOF mass window & $\pm$3$\sigma $ \\
    \hline
  \end{tabular}
  \caption{Selection criteria applied for the $ ^{4} $He and $^{4}\overline{\mathrm{He}}$ analyses.}
  \label{tab:cuts}
\end{table}

%For the identification of the (anti-)$^4$He candidates the TPC and the TOF detector are used. \\
The $\mathrm{d}E/\mathrm{d}x $ is measured in the TPC as a function of the rigidity $ p/z $, where $ p $ is the momentum and $ z $ is the electric charge in units of the elementary charge $ e $. This distribution of reconstructed charged particles is well described by the Bethe-Bloch formula~\cite{bethe,bloch} and is unique for each particle species.  %\textcolor{red}{The energy deposit resolution of the TPC  in central Pb--Pb collisions investigated here is typically around 7\%. }%The particle identification pattern observed is demonstrated in Fig.~\ref{fig:dEdx}.} 

Primarily, all events with at least one particle with a $ \mathrm{d}E/\mathrm{d}x $ corresponding to a $ ^{3} $He and $^{3}\overline{\mathrm{He}}$ or a higher mass are selected. To ensure a good track matching between the TPC and the TOF detectors, only candidates within 3 standard deviations ($\sigma$) around the mean in the $\mathrm{d}E/\mathrm{d}x $ (TPC) vs. $ \beta \gamma $ (TOF) plane are accepted. Here, $\beta$ denotes the relativistic velocity $\beta=v/c$ and $\gamma$ is the Lorentz factor. In order to select $^4$He or $^{4}\overline{\mathrm{He}}$ particles, candidates within a 3$\sigma$ band of the Bethe-Bloch parametrisation in the $\mathrm{d}E/\mathrm{d}x $ versus $ p/z $ distribution are taken into account. At higher momenta, the two Bethe-Bloch curves of $ ^{4} $He or $^{4}\overline{\mathrm{He}}$ and of $ ^{3} $He or $^{3}\overline{\mathrm{He}}$ approach each other. To study a possible contamination from $  ^{3}$He and $^{3}\overline{\mathrm{He}}$ particles, different narrower cuts for the TPC $\mathrm{d}E/\mathrm{d}x $ selection band are investigated: while the upper cut of the band (3$\sigma$) is fixed, the lower cut is restricted progressively going in steps of 0.5 units from -3$\sigma$ up to 0$\sigma$.  For all these seven cuts the procedure described in the following is carried out and a yield d$ N $/d$ y $ is determined.
% Starting from -3$\sigma$ to 3$\sigma$ the band was restricted more and more in 0.5er steps going to -2.5, -2, -1.5 and so on, up to 0$\sigma$ to 3$\sigma$. The higher cut on 3$\sigma$ is always kept. For all these seven cuts the following described procedure has been carried out and a yield d$ N $/d$ y $ has been determined.

\begin{figure}[h!]
	\centering
	\includegraphics[width=\textwidth]{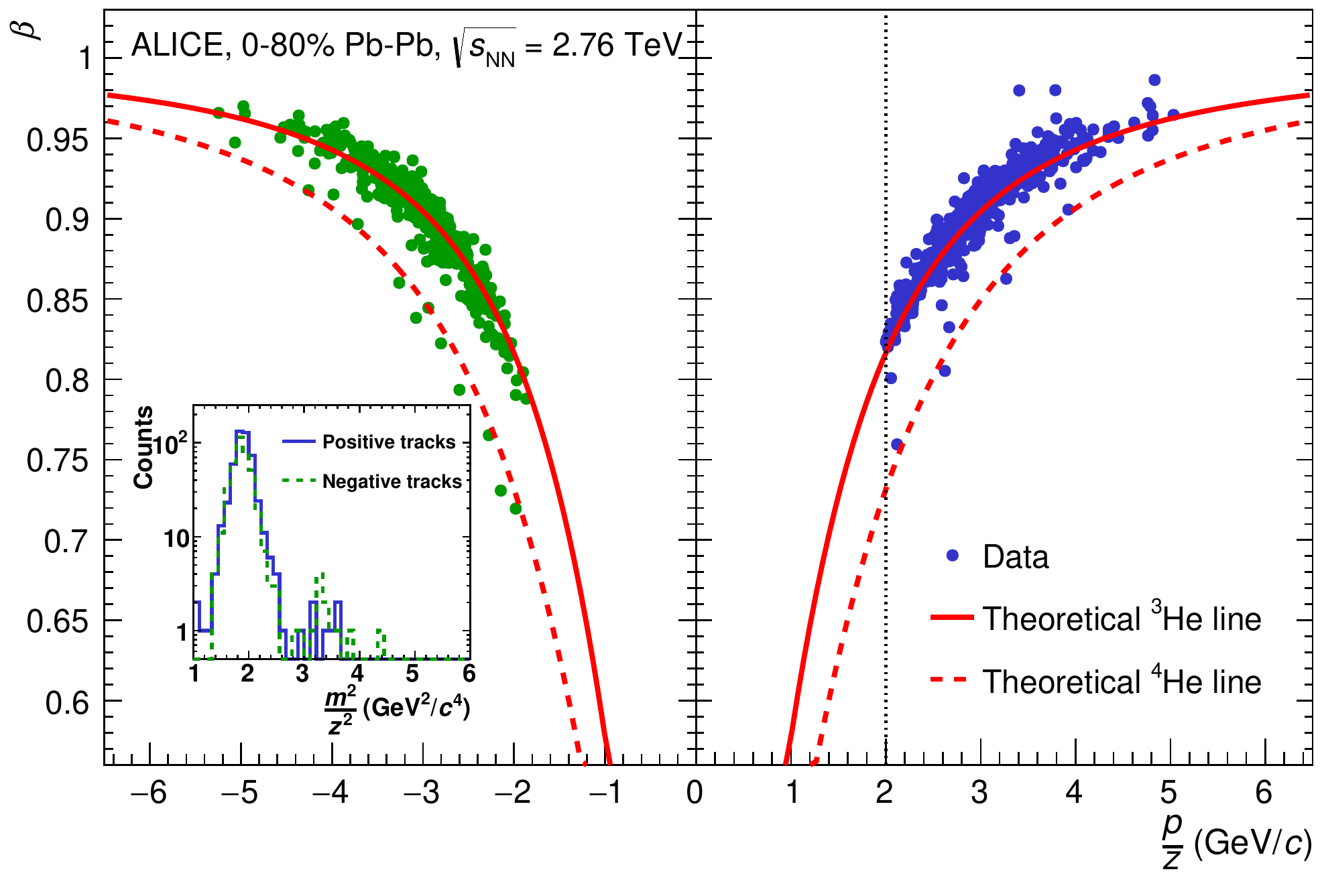} 
	\caption{Velocity $ \beta $ measured with the TOF detector as a function of the rigidity $ p/z $. For this figure a selection band of -1.5 to 3$\sigma$ around the mean of the TPC specific energy-loss distribution is required.
Negatively (positively) charged particles are shown on the left (right) side, with positive tracks in blue and negative tracks in green. The dashed vertical line represents the cut on the rigidity $p/z = 2$~GeV/$c$ (applied only for positively charged particles). The insert shows the $m^2/z^2$ distributions obtained from the data points shown in the main figure.}
	\label{fig:TOF}
\end{figure}

In Figure~\ref{fig:TOF}, the velocity ($\beta$) distributions of He candidates are plotted versus rigidity. One can clearly see the separation of $^3$He and $^4$He. From these data, the $m^{2}/z^{2}$ ($ m $ = mass of the particle) distributions are calculated and displayed in the insert of this figure. From the insert, the separation of $^3$He and $^4$He can be quantitatively asserted.
The $ m^{2}/z^{2}$ is different for $ ^{3} $He (2.00 GeV$ ^{2} / c ^{4} $) and $ ^{4} $He (3.48 GeV$ ^{2} / c ^{4} $). Candidates lying within a window of 2.86 GeV$ ^{2} /c ^{4} $ $ <  m^{2}/z^{2} < $ 4.87 GeV$ ^{2} /c ^{4} $ are identified as $^4$He or $^{4}\overline{\mathrm{He}}$ particles. This window is determined by a fit to the peak in the $ m^{2}/z^{2} $ distribution of the selected tracks. Because of the low statistics, the fitting is done simultaneously both  for particles and for anti-particles, including secondary $^4$He knocked out from the material. A Gaussian with an exponential tail on the right side is used as the fit function. For the background, the sum of a first-order polynomial and an exponential shape is assumed. This is necessary to describe the time-signal shape of the TOF detector~\cite{alice_tof}. The polynomial shape is needed to cope with mismatched candidate tracks in the signal region. A similar procedure is used in~\cite{nuclei}.

For the analysis of positively charged $^4$He, contamination from $^4$He nuclei which do not originate from the primary vertex, but stem from the detector material due to knockout processes, are taken into account. Monte Carlo studies suggest a cut on $ p/z > 2 $ GeV/$c$ to eliminate such a background. Note that the background due to knockout processes is steeply falling with momentum and the signal is rising in this momentum range. Therefore, only $^4$He candidates with a $ p/z $ greater than 2 GeV/$ c $ are accepted. The contamination at higher momenta is estimated to be a maximum of 0.13 counts out of a total count of the order of 10, which is added as a systematic uncertainty.

%In order to estimate the background underneath the $^4$He/$^{4}\overline{\mathrm{He}}$ peak in the TOF mass window, which originates from mismatched tracks in the TOF detector a likelihood fit, with a constant function, outside the peak region was performed. This was done choosing candidates outside the 3$\sigma$ band in the $\mathrm{d}E/\mathrm{d}x $ vs.~$ \beta \gamma $ plane.

%In order to estimate the background underneath the $^4$He/$^{4}\overline{\mathrm{He}}$ peak in the TOF mass window, which originates from mismatched tracks in the TOF detector, a likelihood fit under the assumption of a flat background is done in the $\mathrm{d}E/\mathrm{d}x $ vs.~$ \beta \gamma $ plane outside the $\pm 3\sigma$ matching band. In this way these candidates are identified as mismatched particles, which are normally rejected and only used for this purpose. Due to limited statistics only candidates from the window defined by the centre of the distribution to 3$ \sigma $ away from of the Bethe-Bloch parametrisation a different approach is used. For this particular window, we assumed a constant ratio of $ ^{3} $He to background counts (due to TOF mismatch).% $R = ^{3}$He/(Background Counts). 

The small number of clear signal counts observed by combining the TPC and TOF information does not give any indication of background. In order to estimate an upper limit on the background counts from mismatched tracks in the TOF detector underneath the $^4$He or $^{4}\overline{\mathrm{He}}$ peak in the TOF mass window, a likelihood fit under the assumption of a flat background is performed in the $\mathrm{d}E/\mathrm{d}x $ versus~$ \beta \gamma $ plane outside the $\pm 3\sigma$ matching band. In this way, background candidates are identified as mismatched particles. (These are usually rejected and only used for this purpose.) Due to limited statistics, this procedure cannot be used if a stronger selection criterion is applied for the TPC $\mathrm{d}E/\mathrm{d}x$ selection, since no $^4$He or $^{4}\overline{\mathrm{He}}$ candidates are left to apply this technique. For these particular cases, we assume a constant ratio of $^{3}$He to background counts and use this to estimate the number of $^{4}$He background.

The background stemming from misidentification of (anti-)$^3$He as (anti-)$^4$He is estimated to be more than one order of magnitude smaller than the one from the mismatch of TPC tracks when extrapolated to the TOF detector and is therefore considered to be negligible. The estimated background decreases with more stringent TPC $\mathrm{d}E/\mathrm{d}x $ cuts. The signal-to-background ratio improves depending on the tightness of the  $\mathrm{d}E/\mathrm{d}x $ cut from 1.7 to 8.4 for $^4$He and from 1.7 to 17.6 for $^{4}\overline{\mathrm{He}}$.

To estimate the efficiency for the detection of $^4$He and $^{4}\overline{\mathrm{He}}$, a Monte Carlo simulation is generated in which the kinematical distributions of the particles are generated flat both in rapidity $y$ and in transverse momentum $p_{\mathrm{T}}$. The shape of $p_{\mathrm{T}}$ spectra in heavy-ion collisions is typically described by a blast-wave model~\cite{Schnedermann:1993ws}. This model assumes an average radial-flow velocity $ \langle \beta \rangle$ and a kinetic freeze-out temperature $T_\mathrm{kin}$ as described in~\cite{pKpi}. Generally, most hadron $p_{\mathrm{T}}$ spectra measured in heavy-ion collisions can be described well by one common set of parameters~\cite{pKpi_centrality}. Surprisingly, this also works well for the description of deuteron and $ ^{3} $He $p_{\mathrm{T}}$ spectra~\citep{nuclei}. Hence the same prescription is used here for the $p_{\mathrm{T}}$ shape of $^4$He and $^{4}\overline{\mathrm{He}}$ particles, namely the same set of parameters is used, only the mass is changed to the $^4$He mass. 

Since only a small number of $^4$He and $^{4}\overline{\mathrm{He}}$ particles (14 $^{4}\overline{\mathrm{He}}$ and 9 $^4$He for the widest TPC $\mathrm{d}E/\mathrm{d}x $ cut) are observed, a $ p_{\mathrm{T}}$ spectrum can not be measured. It is estimated using the blast-wave parameters of deuterons and $ ^{3}\mathrm{He}$ spectra~\cite{nuclei}. The final acceptance $ \times $ efficiencies are obtained as described in~\cite{exotica_paper} and are of the order of  15\% for $^4$He and 20\% for $^{4}\overline{\mathrm{He}}$. The difference originates from the 2 GeV/$ c $ rigidity cut applied to $^4$He candidates.

For the $^{4}\overline{\mathrm{He}}$ analysis, the absorption in the detector material is taken into account using two different transport codes, namely GEANT3~\cite{GEANT3} and GEANT4~\cite{geant4}. These two codes use different models for the estimation of the absorption cross section. In GEANT4, a Glauber model based on the well known hadronic interaction cross sections for (anti-)protons is implemented~\cite{geant4model}. The version of GEANT3 used in this analysis is modified~\cite{nuclei} such that it calculates the absorption based on an empirical parameterisation~\cite{eulogio}, based on the measurements of anti-deuterons carried out at Serpukhov~\cite{serpukhov}. The baseline is given by the absorption calculated with GEANT4, while the GEANT3 based correction is used in the systematic uncertainty evaluation. The maximum absorption probability towards low $ p/z $ is about 20\%. In contrast to GEANT4, which still shows an absorption of about 5\% at $p_{\mathrm{T}} = 10$~GeV/$c$, GEANT3 exhibits basically no absorption above 3.5 GeV/$c$. 

The main contributions to the systematic uncertainty on the determined production yields are: 

\begin{itemize}

\item The uncertainty due to the unknown shape of the $ p_{\mathrm{T}}$ distributions, which is determined by using the blast-wave model based on the measured deuteron and $^{3}\mathrm{He}$ spectra~\cite{nuclei}. This leads to a systematic uncertainty contribution of around 13\%.

\item Only for $^4$He: The rigidity cut on $ p/z $ greater than 2 GeV/$ c $ itself has a systematic uncertainty of 4 to 13\% depending on the TPC PID cut. As mentioned before, the  secondary contamination above this cut is estimated to be a maximum of 0.13 counts. This leads to a systematic uncertainty of at minimum 20\% and at maximum 49\%  growing with stricter TPC PID cut. As the number of observed candidates shrinks with stricter TPC $\mathrm{d}E/\mathrm{d}x $ selection, the systematic uncertainty on the secondary contamination grows.

\item Only for $^{4}\overline{\mathrm{He}}$: The absorption correction has an uncertainty of 7\%, estimated from the difference of the two GEANT implementations.

\end{itemize} 

Other systematic uncertainties are estimated by varying the cuts in the limits consistent with the detector resolution. The contributions of these systematic uncertainties are typically found to be below the percent range. The systematic uncertainty on the chosen TPC PID cut varies between 1\% for the most loose cuts and 19\% for stricter cuts. This is caused by the stronger sensitivity of the stricter cuts, namely the even further reduced low number of candidates, which is not reflected in the Monte Carlo simulation.

The final values and the corresponding uncertainties are calculated as a mean from the previously discussed variations of the selection criteria. The resulting systematic uncertainty on the final yield is 35\% for $^4$He and 20\% for $^{4}\overline{\mathrm{He}}$.
 
%are varying strongly between 1 and 30\%, depending on the lower n$ \sigma $ value. This has to do with the small statistics for the stricter cuts.

%\begin{figure}
%	\centering
%	\includegraphics[width=0.49\textwidth]{dEdx_pos} 
%	\includegraphics[width=0.49\textwidth]{dEdx_neg} 
%	\caption{TPC $ \mathrm{d}E /\mathrm{d}x $ spectrum for positive (left) and negative (right) particles. The solid lines are parametrisations of the Bethe-Bloch-formula~\cite{bethe,bloch,aleph} to describe the different particle species. The dashed line indicates the "offline" trigger to select all events, which contain a (anti-)$ ^{3} $He or a higher mass.}
%	\label{fig:dEdx}
%\end{figure}
 
\section{Results}

The measurement is performed on a data set including central, semi-central and minimum-bias triggered events. To make use of all the data analysed, the semi-central and minimum-bias events have been extrapolated to 0-10\% centrality interval assuming that the particle and anti-particle yields scale linearly with the charged-particle multiplicity d$ N_{\mathrm{ch}} $/d$ \eta $. This procedure has already been tested to work well for the (anti-)hypertriton production~\cite{hypertriton}. In addition, d/p and $^3$He/p ratios are measured to be approximately flat versus multiplicity within uncertainties\cite{nuclei}. Thus, for each centrality class, the number of analysed events is multiplied by the corresponding measured charged-particle density d$ N_{\mathrm{ch}} $/d$ \eta $~\cite{vzero2}. If this is added up and divided by the total number of measured events it leads to a weighting factor of 1034. To get the final yield in the 0-10\% centrality class the measured yield is multiplied with the d$ N_{\mathrm{ch}} $/d$ \eta $ for 0-10\% centrality (1447.5) and divided by the weighting factor, as $ \mathrm{d}N / \mathrm{d}y_{0-10\%} = \mathrm{d}N / \mathrm{d}y_{measured} \times 1447.5 / 1034 $. 

%\begin{equation}
%\frac{\mathrm{d}N}{\mathrm{d}y}_{0-10\%} = \frac{\mathrm{d}N}{\mathrm{d}y}_{measured} \times \frac{ 2.072 \cdot 10^{7} \times 1447.5 + 1.740 \cdot 10^{7} \times 749 + 0.063 \cdot 10^{7} \times 86 } {3.874\cdot 10^{7}} \times 1447.5
%\end{equation}

%\begin{equation}
%\frac{\mathrm{d}N}{\mathrm{d}y}_{0-10\%} = \frac{\mathrm{d}N}{\mathrm{d}y}_{measured} \times \frac{1447.5}{1112}
%\end{equation}

%This leads to final values of d$ N $/d$ y $ = $ 6.8 \times 10^{-7} \pm 3.2 \times 10^{-7}  (\mathrm{stat}) \pm 3.5 \times 10^{-7}  (\mathrm{syst})  $ for $^4$He and d$ N $/d$ y $ = $ 9.8 \times 10^{-7} \pm 3.4 \times 10^{-7}  (\mathrm{stat}) \pm 1.6 \times 10^{-7}  (\mathrm{syst})  $ for $^{4}\overline{\mathrm{He}}$. For the ratio $^{4}\overline{\mathrm{He}}/^{4}\mathrm{He}$  we obtain $1.4 \pm 0.8 (\mathrm{stat}) \pm 0.8 (\mathrm{syst})$. ("stat" and "syst" indicate the statistical and the systematic uncertainty.) \\%The ratio $ \overline{^{4}\mathrm{He}}/^{4}\mathrm{He}$  is $1.44 \pm 0.83 (\mathrm{stat.}) \pm 0.68 (\mathrm{sys.})$. ("stat." and "sys." indicate the statistical and the systematic uncertainty.) \\

This leads to final values of  $\mathrm{d}N/\mathrm{d}y _{^{4}\mathrm{He}} =  (0.8  \pm 0.4 ~(\mathrm{stat}) \pm 0.3~(\mathrm{syst}))\times 10^{-6}$ for $^4$He and $\mathrm{d}N/\mathrm{d}y _{^{4}\mathrm{\overline{He}}} =  (1.1  \pm 0.4~(\mathrm{stat}) \pm 0.2~(\mathrm{syst}))\times 10^{-6}$ for $^{4}\overline{\mathrm{He}}$. For the ratio $^{4}\overline{\mathrm{He}}/^{4}\mathrm{He}$  we obtain $1.4 \pm 0.8 (\mathrm{stat}) \pm 0.5 (\mathrm{syst})$ ("stat" and "syst" indicate the statistical and the systematic uncertainty).

\begin{figure}[h!]
	\centering
	\includegraphics[width=0.8\textwidth]{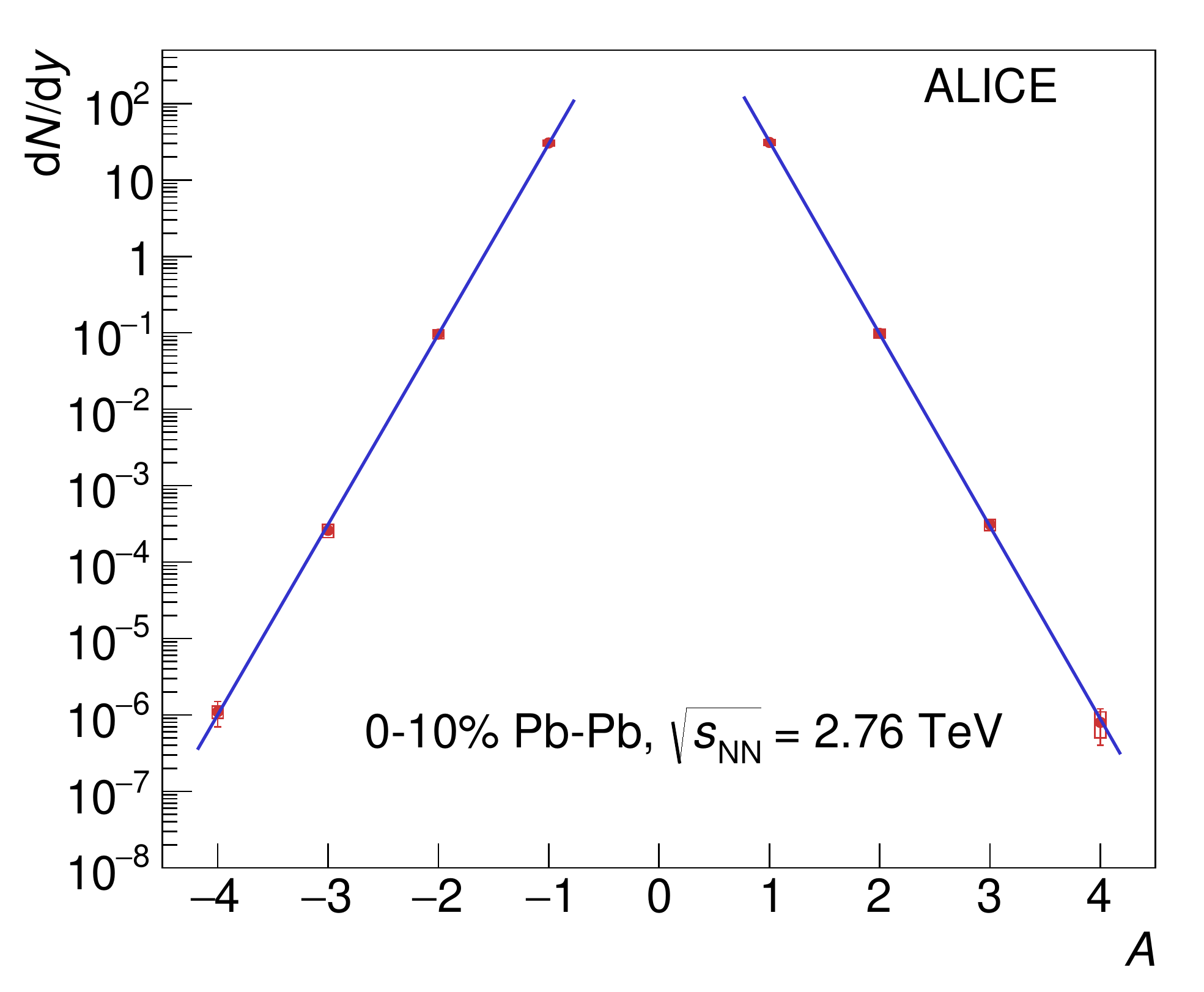}
	\caption{d$ N $/d$ y $ for protons (A=1) up to $^4$He (A=4) and the corresponding anti-particles in central (0-10\%) Pb--Pb collisions at $\sqrt{s_{\mathrm{NN}}}$ = 2.76 TeV. The blue lines are fits with an exponential function. Statistical uncertainties are shown as lines, whereas the systematic uncertainties are represented by boxes.}
		\label{fig:MassOrdering}
\end{figure}

The measured yields in the 0-10\% centrality interval are shown in Fig.~\ref{fig:MassOrdering} together with those of \makebox{(anti-)protons}, (anti-)deuterons and (anti-)$^{3} $He~\cite{nuclei,pKpi_centrality} (details on the extrapolation to 0-10\% centrality can be found in~\cite{QM2014Michele}). The blue lines are exponential fits with the fit function $ Ke^{BA} $ resulting in $ B = -5.8 \pm 0.2 $, which corresponds to a penalty factor (suppression factor of production yield for nuclei with one additional baryon) of around 300. The same penalty factor is also obtained if the fit is done up to $ ^{3}$He only~\cite{nuclei}.
 
%\section{Discussion}

\begin{figure}[h!]
	\centering
	\includegraphics[width=0.99\textwidth]{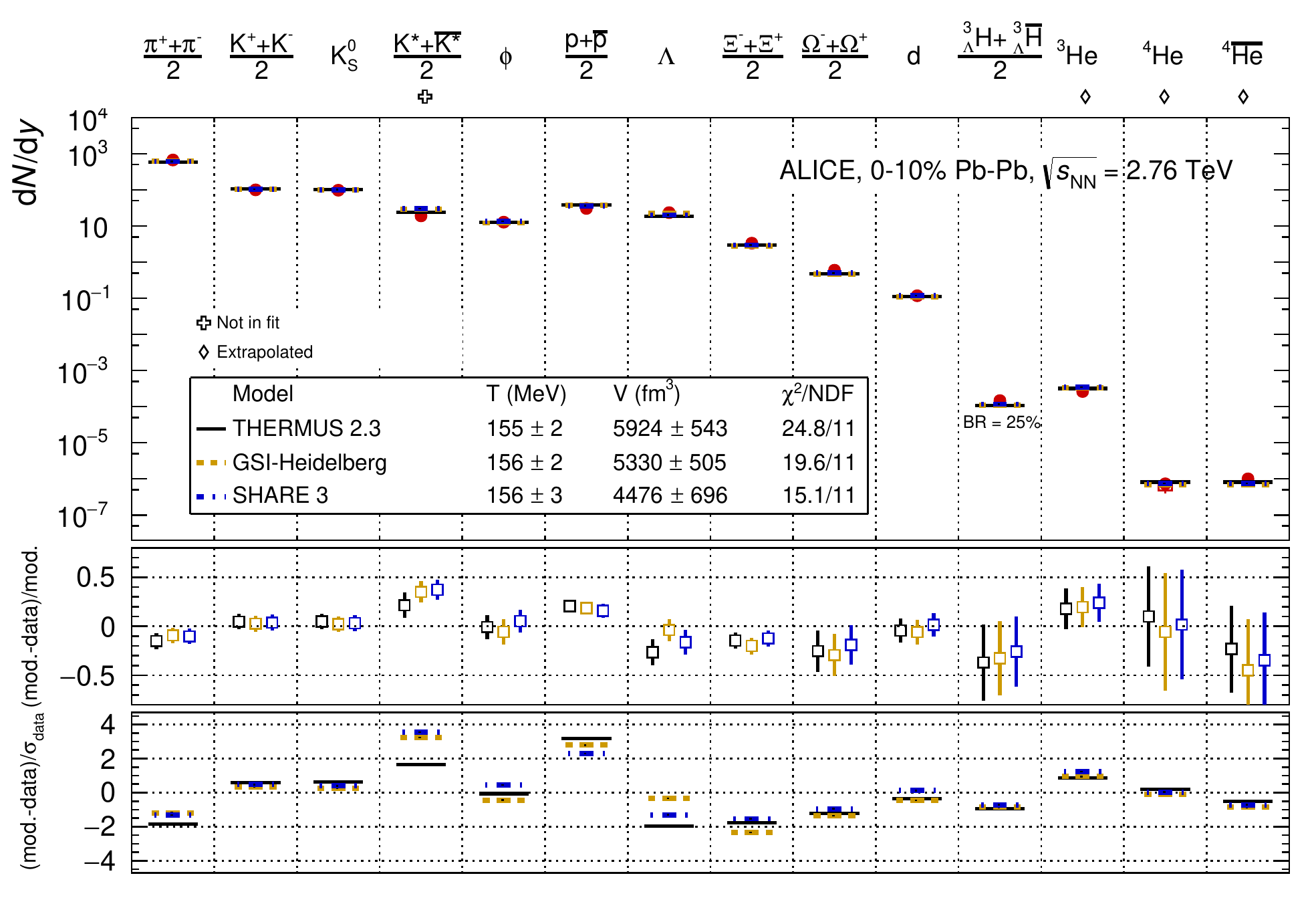}
	\caption{Thermal model fits, with three different implementations, to the light flavour hadron yields in central (0-10\%) Pb--Pb collisions at $\sqrt{s_{\mathrm{NN}}}$ = 2.76 TeV. The data points are taken from~\cite{pKpi_centrality,lambda_k0s,MultiStrange,nuclei,hypertriton,Phi_K,Adam:2017zbf} and details of the fits can be found in~\cite{QM2014Michele,Andronic:2016nof}. The upper panel shows the fit results together with the data, whereas the middle panel shows the difference between model and data normalised to the model value and the lower panel the difference between model and data normalised to the experimental uncertainties.}
		\label{fig:ThermalFit}
\end{figure}

The obtained penalty factor of around 300 for each additional nucleon is consistent with $T_{\mathrm{chem}} \approx 160$ MeV in the equilibrium thermal models. The measured yields for $^4$He and $^{4}\overline{\mathrm{He}}$ nuclei are consistent with the predictions  from the various (equilibrium) thermal models (THERMUS~\cite{Thermus}, GSI~\cite{GSI,thermalModel,Andronic:2017pug} and SHARE~\cite{Share1,Share2,Share3}) with $T_{\mathrm{chem}}$ = 156 MeV, as shown in Fig.~\ref{fig:ThermalFit} for complete statistical thermal model fits using the available light flavour data measured by the ALICE Collaboration. The fits in Fig.~\ref{fig:ThermalFit} extend the  simple exponential model (Fig.~\ref{fig:MassOrdering}) by incorporating Boltzmann statistics and degeneracy factors for all particles. If instead of all listed particles only nuclei (deuterons, $^{3}$He and  $^4$He and $^{4}\overline{\mathrm{He}}$) are considered for the fit, the resulting temperatures are 154 $\pm$ 4 MeV. The pure measured yields for $^4$He and $^{4}\overline{\mathrm{He}}$ nuclei agree, depending on the model implementation, within the determined uncertainties with temperatures from 135 MeV to 177 MeV. Taken together these observations suggest that the relatively heavy $^4$He and $^{4}\overline{\mathrm{He}}$ nuclei are also produced statistically at the same temperature as the lighter particles.
%~\footnote{Please note, that for this fits no feed-down has to be considered as the hypertriton for example is suppressed by factor 330.}

\section{Summary and Conclusion}

The ALICE Collaboration has measured the production yields of $^4$He and $^{4}\overline{\mathrm{He}}$ in central (0-10\%) Pb--Pb collisions at $\sqrt{s_{\mathrm{NN}}}$ = 2.76 TeV. The ratio of the two yields is consistent with unity and the results are in good agreement with the prediction of the statistical thermal model assuming the same temperature of 156 MeV as is obtained from the fit to the other light flavour hadrons.

%With the future runs 2 and 3 of the LHC it is expected that ALICE can measure the yields of $^4$He and $^{4}\overline{\mathrm{He}}$ particles with more precision. The higher statistics will allow for a measurement of the $^{4}\overline{\mathrm{He}}$ transverse momentum spectrum. As the unknown shape of the $ p_{\mathrm{T}}$ distributions is one of the major source of the systematic uncertainty the measurement of the spectrum will decrease the systematic uncertainty of the measured yield. Furthermore with higher statistics it will be also possible to measure the absorption of the $^{4}\overline{\mathrm{He}}$, which is currently not possible. Including all this it will be possible to finally answer the question if the ratio $ \overline{^{4}\mathrm{He}}/^{4}\mathrm{He}$ differs from one.

%As visible from the current paper the statistics limits the precision of the measurement strongly.
Data gathered at the current beam energy of $\sqrt{s_{\mathrm{NN}}}$ = 5.02~TeV in Pb--Pb collisions at the LHC ($ Run$ 2) will improve the studies described in this letter thanks to an increase in statistics by a factor of about 3. Based on the pilot measurement presented here, we conclude that a precision study will be possible in the data taking period starting from 2021 ($ Run$ 3 of the LHC), where about 5500 $^4$He($^{4}\overline{\mathrm{He}}$) particles are expected to be reconstructed~\cite{LoI}. This will allow for the measurement of the transverse-momentum spectra. As the unknown shape of the $ p_{\mathrm{T}}$ distributions is one of the major sources of the systematic uncertainty, the measurement of the spectrum will decrease the systematic uncertainty of the measured yield. As a consequence the precision of the  ratio of $^{4}\overline{\mathrm{He}}/^{4}\mathrm{He}$ will be significantly improved. In addition, a mass difference measurement similar to what was done in~\cite{MassDiff} will be possible.
               %%%%%%%%%%% put the body of the article here
%
%

%%%%% acknowledgements
\newenvironment{acknowledgement}{\relax}{\relax}
\begin{acknowledgement}
\section*{Acknowledgements}
%We thank bla and blub for useful correspondence.
% Version: 2017-10-17

The ALICE Collaboration would like to thank all its engineers and technicians for their invaluable contributions to the construction of the experiment and the CERN accelerator teams for the outstanding performance of the LHC complex.
The ALICE Collaboration gratefully acknowledges the resources and support provided by all Grid centres and the Worldwide LHC Computing Grid (WLCG) collaboration.
The ALICE Collaboration acknowledges the following funding agencies for their support in building and running the ALICE detector:
A. I. Alikhanyan National Science Laboratory (Yerevan Physics Institute) Foundation (ANSL), State Committee of Science and World Federation of Scientists (WFS), Armenia;
Austrian Academy of Sciences and Nationalstiftung f\"{u}r Forschung, Technologie und Entwicklung, Austria;
Ministry of Communications and High Technologies, National Nuclear Research Center, Azerbaijan;
Conselho Nacional de Desenvolvimento Cient\'{\i}fico e Tecnol\'{o}gico (CNPq), Universidade Federal do Rio Grande do Sul (UFRGS), Financiadora de Estudos e Projetos (Finep) and Funda\c{c}\~{a}o de Amparo \`{a} Pesquisa do Estado de S\~{a}o Paulo (FAPESP), Brazil;
Ministry of Science \& Technology of China (MSTC), National Natural Science Foundation of China (NSFC) and Ministry of Education of China (MOEC) , China;
Ministry of Science, Education and Sport and Croatian Science Foundation, Croatia;
Ministry of Education, Youth and Sports of the Czech Republic, Czech Republic;
The Danish Council for Independent Research | Natural Sciences, the Carlsberg Foundation and Danish National Research Foundation (DNRF), Denmark;
Helsinki Institute of Physics (HIP), Finland;
Commissariat \`{a} l'Energie Atomique (CEA) and Institut National de Physique Nucl\'{e}aire et de Physique des Particules (IN2P3) and Centre National de la Recherche Scientifique (CNRS), France;
Bundesministerium f\"{u}r Bildung, Wissenschaft, Forschung und Technologie (BMBF) and GSI Helmholtzzentrum f\"{u}r Schwerionenforschung GmbH, Germany;
General Secretariat for Research and Technology, Ministry of Education, Research and Religions, Greece;
National Research, Development and Innovation Office, Hungary;
Department of Atomic Energy Government of India (DAE), Department of Science and Technology, Government of India (DST), University Grants Commission, Government of India (UGC) and Council of Scientific and Industrial Research (CSIR), India;
Indonesian Institute of Science, Indonesia;
Centro Fermi - Museo Storico della Fisica e Centro Studi e Ricerche Enrico Fermi and Istituto Nazionale di Fisica Nucleare (INFN), Italy;
Institute for Innovative Science and Technology , Nagasaki Institute of Applied Science (IIST), Japan Society for the Promotion of Science (JSPS) KAKENHI and Japanese Ministry of Education, Culture, Sports, Science and Technology (MEXT), Japan;
Consejo Nacional de Ciencia (CONACYT) y Tecnolog\'{i}a, through Fondo de Cooperaci\'{o}n Internacional en Ciencia y Tecnolog\'{i}a (FONCICYT) and Direcci\'{o}n General de Asuntos del Personal Academico (DGAPA), Mexico;
Nederlandse Organisatie voor Wetenschappelijk Onderzoek (NWO), Netherlands;
The Research Council of Norway, Norway;
Commission on Science and Technology for Sustainable Development in the South (COMSATS), Pakistan;
Pontificia Universidad Cat\'{o}lica del Per\'{u}, Peru;
Ministry of Science and Higher Education and National Science Centre, Poland;
Korea Institute of Science and Technology Information and National Research Foundation of Korea (NRF), Republic of Korea;
Ministry of Education and Scientific Research, Institute of Atomic Physics and Romanian National Agency for Science, Technology and Innovation, Romania;
Joint Institute for Nuclear Research (JINR), Ministry of Education and Science of the Russian Federation and National Research Centre Kurchatov Institute, Russia;
Ministry of Education, Science, Research and Sport of the Slovak Republic, Slovakia;
National Research Foundation of South Africa, South Africa;
Centro de Aplicaciones Tecnol\'{o}gicas y Desarrollo Nuclear (CEADEN), Cubaenerg\'{\i}a, Cuba, Ministerio de Ciencia e Innovacion and Centro de Investigaciones Energ\'{e}ticas, Medioambientales y Tecnol\'{o}gicas (CIEMAT), Spain;
Swedish Research Council (VR) and Knut \& Alice Wallenberg Foundation (KAW), Sweden;
European Organization for Nuclear Research, Switzerland;
National Science and Technology Development Agency (NSDTA), Suranaree University of Technology (SUT) and Office of the Higher Education Commission under NRU project of Thailand, Thailand;
Turkish Atomic Energy Agency (TAEK), Turkey;
National Academy of  Sciences of Ukraine, Ukraine;
Science and Technology Facilities Council (STFC), United Kingdom;
National Science Foundation of the United States of America (NSF) and United States Department of Energy, Office of Nuclear Physics (DOE NP), United States of America.    %%%%%%% get the lates version before submitting
\end{acknowledgement}

%%%%%%%% Bibliography (In case of using bibtex generate the bbl requested by arXiv)
\bibliographystyle{utphys}   % Put here the style file name for the paper, i.e.apsrev4-1, utphys
%\bibliographystyle{ieeetr}   % Put here the style file name for the paper, i.e.apsrev4-1, utphys
%\bibliography{biblio}
\bibliography{alpha_paper}
\newpage
%
%\input{}               %%%%%%%%%%% put your appendices here
%
%%%%%%%%% appendix with author list
\appendix
\section{The ALICE Collaboration}
\label{app:collab}
% Collaboration: CERN-LHC-ALICE
% Generation Date is 2017-Sep-13

% How to use:
%%%%%%%%% appendix with author list
%\appendix
%\section{The ALICE Collaboration}
%\label{app:collab}
%\input{Alice_Authorslist_XXXX-Axx-XX.tex}
\begingroup
\small
\begin{flushleft}
S.~Acharya\Irefn{org137}\And 
D.~Adamov\'{a}\Irefn{org94}\And 
J.~Adolfsson\Irefn{org34}\And 
M.M.~Aggarwal\Irefn{org99}\And 
G.~Aglieri Rinella\Irefn{org35}\And 
M.~Agnello\Irefn{org31}\And 
N.~Agrawal\Irefn{org48}\And 
Z.~Ahammed\Irefn{org137}\And 
S.U.~Ahn\Irefn{org79}\And 
S.~Aiola\Irefn{org141}
% ...
\And
\renewcommand\labelenumi{\textsuperscript{\theenumi}~}

\section*{Affiliation notes}
\renewcommand\theenumi{\roman{enumi}}
\begin{Authlist}
\item \Adef{org*}Deceased
\item \Adef{orgI}Dipartimento DET del Politecnico di Torino, Turin, Italy
\item \Adef{orgII}M.V. Lomonosov Moscow State University, D.V. Skobeltsyn Institute of Nuclear, Physics, Moscow, Russia
\item \Adef{orgIII}Department of Applied Physics, Aligarh Muslim University, Aligarh, India
\item \Adef{orgIV}Institute of Theoretical Physics, University of Wroclaw, Poland
\end{Authlist}

\section*{Collaboration Institutes}
\renewcommand\theenumi{\arabic{enumi}~}
\begin{Authlist}
\item \Idef{org1}A.I. Alikhanyan National Science Laboratory (Yerevan Physics Institute) Foundation, Yerevan, Armenia
\item \Idef{org2}Benem\'{e}rita Universidad Aut\'{o}noma de Puebla, Puebla, Mexico
\item \Idef{org3}Bogolyubov Institute for Theoretical Physics, Kiev, Ukraine
\item \Idef{org4}Bose Institute, Department of Physics  and Centre for Astroparticle Physics and Space Science (CAPSS), Kolkata, India
\item \Idef{org5}Budker Institute for Nuclear Physics, Novosibirsk, Russia
\item \Idef{org6}California Polytechnic State University, San Luis Obispo, California, United States
\item \Idef{org7}Central China Normal University, Wuhan, China
\item \Idef{org8}Centre de Calcul de l'IN2P3, Villeurbanne, Lyon, France
\item \Idef{org9}Centro de Aplicaciones Tecnol\'{o}gicas y Desarrollo Nuclear (CEADEN), Havana, Cuba
\item \Idef{org10}Centro de Investigaciones Energ\'{e}ticas Medioambientales y Tecnol\'{o}gicas (CIEMAT), Madrid, Spain
\item \Idef{org11}Centro de Investigaci\'{o}n y de Estudios Avanzados (CINVESTAV), Mexico City and M\'{e}rida, Mexico
\item \Idef{org12}Centro Fermi - Museo Storico della Fisica e Centro Studi e Ricerche ``Enrico Fermi', Rome, Italy
\item \Idef{org13}Chicago State University, Chicago, Illinois, United States
\item \Idef{org14}China Institute of Atomic Energy, Beijing, China
\item \Idef{org15}COMSATS Institute of Information Technology (CIIT), Islamabad, Pakistan
\item \Idef{org16}Departamento de F\'{\i}sica de Part\'{\i}culas and IGFAE, Universidad de Santiago de Compostela, Santiago de Compostela, Spain
\item \Idef{org17}Department of Physics, Aligarh Muslim University, Aligarh, India
\item \Idef{org18}Department of Physics, Ohio State University, Columbus, Ohio, United States
\item \Idef{org19}Department of Physics, Pusan National University, Pusan, Republic of Korea
\item \Idef{org20}Department of Physics, Sejong University, Seoul, Republic of Korea
\item \Idef{org21}Department of Physics, University of Oslo, Oslo, Norway
\item \Idef{org22}Department of Physics and Technology, University of Bergen, Bergen, Norway
\item \Idef{org23}Dipartimento di Fisica dell'Universit\`{a} 'La Sapienza' and Sezione INFN, Rome, Italy
\item \Idef{org24}Dipartimento di Fisica dell'Universit\`{a} and Sezione INFN, Cagliari, Italy
\item \Idef{org25}Dipartimento di Fisica dell'Universit\`{a} and Sezione INFN, Trieste, Italy
\item \Idef{org26}Dipartimento di Fisica dell'Universit\`{a} and Sezione INFN, Turin, Italy
\item \Idef{org27}Dipartimento di Fisica e Astronomia dell'Universit\`{a} and Sezione INFN, Bologna, Italy
\item \Idef{org28}Dipartimento di Fisica e Astronomia dell'Universit\`{a} and Sezione INFN, Catania, Italy
\item \Idef{org29}Dipartimento di Fisica e Astronomia dell'Universit\`{a} and Sezione INFN, Padova, Italy
\item \Idef{org30}Dipartimento di Fisica `E.R.~Caianiello' dell'Universit\`{a} and Gruppo Collegato INFN, Salerno, Italy
\item \Idef{org31}Dipartimento DISAT del Politecnico and Sezione INFN, Turin, Italy
\item \Idef{org32}Dipartimento di Scienze e Innovazione Tecnologica dell'Universit\`{a} del Piemonte Orientale and INFN Sezione di Torino, Alessandria, Italy
\item \Idef{org33}Dipartimento Interateneo di Fisica `M.~Merlin' and Sezione INFN, Bari, Italy
\item \Idef{org34}Division of Experimental High Energy Physics, University of Lund, Lund, Sweden
\item \Idef{org35}European Organization for Nuclear Research (CERN), Geneva, Switzerland
\item \Idef{org36}Excellence Cluster Universe, Technische Universit\"{a}t M\"{u}nchen, Munich, Germany
\item \Idef{org37}Faculty of Engineering, Bergen University College, Bergen, Norway
\item \Idef{org38}Faculty of Mathematics, Physics and Informatics, Comenius University, Bratislava, Slovakia
\item \Idef{org39}Faculty of Nuclear Sciences and Physical Engineering, Czech Technical University in Prague, Prague, Czech Republic
\item \Idef{org40}Faculty of Science, P.J.~\v{S}af\'{a}rik University, Ko\v{s}ice, Slovakia
\item \Idef{org41}Faculty of Technology, Buskerud and Vestfold University College, Tonsberg, Norway
\item \Idef{org42}Frankfurt Institute for Advanced Studies, Johann Wolfgang Goethe-Universit\"{a}t Frankfurt, Frankfurt, Germany
\item \Idef{org43}Gangneung-Wonju National University, Gangneung, Republic of Korea
\item \Idef{org44}Gauhati University, Department of Physics, Guwahati, India
\item \Idef{org45}Helmholtz-Institut f\"{u}r Strahlen- und Kernphysik, Rheinische Friedrich-Wilhelms-Universit\"{a}t Bonn, Bonn, Germany
\item \Idef{org46}Helsinki Institute of Physics (HIP), Helsinki, Finland
\item \Idef{org47}Hiroshima University, Hiroshima, Japan
\item \Idef{org48}Indian Institute of Technology Bombay (IIT), Mumbai, India
\item \Idef{org49}Indian Institute of Technology Indore, Indore, India
\item \Idef{org50}Indonesian Institute of Sciences, Jakarta, Indonesia
\item \Idef{org51}INFN, Laboratori Nazionali di Frascati, Frascati, Italy
\item \Idef{org52}INFN, Sezione di Bari, Bari, Italy
\item \Idef{org53}INFN, Sezione di Bologna, Bologna, Italy
\item \Idef{org54}INFN, Sezione di Cagliari, Cagliari, Italy
\item \Idef{org55}INFN, Sezione di Catania, Catania, Italy
\item \Idef{org56}INFN, Sezione di Padova, Padova, Italy
\item \Idef{org57}INFN, Sezione di Roma, Rome, Italy
\item \Idef{org58}INFN, Sezione di Torino, Turin, Italy
\item \Idef{org59}INFN, Sezione di Trieste, Trieste, Italy
\item \Idef{org60}Inha University, Incheon, Republic of Korea
\item \Idef{org61}Institut de Physique Nucl\'eaire d'Orsay (IPNO), Universit\'e Paris-Sud, CNRS-IN2P3, Orsay, France
\item \Idef{org62}Institute for Nuclear Research, Academy of Sciences, Moscow, Russia
\item \Idef{org63}Institute for Subatomic Physics of Utrecht University, Utrecht, Netherlands
\item \Idef{org64}Institute for Theoretical and Experimental Physics, Moscow, Russia
\item \Idef{org65}Institute of Experimental Physics, Slovak Academy of Sciences, Ko\v{s}ice, Slovakia
\item \Idef{org66}Institute of Physics, Academy of Sciences of the Czech Republic, Prague, Czech Republic
\item \Idef{org67}Institute of Physics, Bhubaneswar, India
\item \Idef{org68}Institute of Space Science (ISS), Bucharest, Romania
\item \Idef{org69}Institut f\"{u}r Informatik, Johann Wolfgang Goethe-Universit\"{a}t Frankfurt, Frankfurt, Germany
\item \Idef{org70}Institut f\"{u}r Kernphysik, Johann Wolfgang Goethe-Universit\"{a}t Frankfurt, Frankfurt, Germany
\item \Idef{org71}Institut f\"{u}r Kernphysik, Westf\"{a}lische Wilhelms-Universit\"{a}t M\"{u}nster, M\"{u}nster, Germany
\item \Idef{org72}Instituto de Ciencias Nucleares, Universidad Nacional Aut\'{o}noma de M\'{e}xico, Mexico City, Mexico
\item \Idef{org73}Instituto de F\'{i}sica, Universidade Federal do Rio Grande do Sul (UFRGS), Porto Alegre, Brazil
\item \Idef{org74}Instituto de F\'{\i}sica, Universidad Nacional Aut\'{o}noma de M\'{e}xico, Mexico City, Mexico
\item \Idef{org75}IRFU, CEA, Universit\'{e} Paris-Saclay, Saclay, France
\item \Idef{org76}iThemba LABS, National Research Foundation, Somerset West, South Africa
\item \Idef{org77}Joint Institute for Nuclear Research (JINR), Dubna, Russia
\item \Idef{org78}Konkuk University, Seoul, Republic of Korea
\item \Idef{org79}Korea Institute of Science and Technology Information, Daejeon, Republic of Korea
\item \Idef{org80}KTO Karatay University, Konya, Turkey
\item \Idef{org81}Laboratoire de Physique Subatomique et de Cosmologie, Universit\'{e} Grenoble-Alpes, CNRS-IN2P3, Grenoble, France
\item \Idef{org82}Lawrence Berkeley National Laboratory, Berkeley, California, United States
\item \Idef{org83}Moscow Engineering Physics Institute, Moscow, Russia
\item \Idef{org84}Nagasaki Institute of Applied Science, Nagasaki, Japan
\item \Idef{org85}National and Kapodistrian University of Athens, Physics Department, Athens, Greece
\item \Idef{org86}National Centre for Nuclear Studies, Warsaw, Poland
\item \Idef{org87}National Institute for Physics and Nuclear Engineering, Bucharest, Romania
\item \Idef{org88}National Institute of Science Education and Research, HBNI, Jatni, India
\item \Idef{org89}National Nuclear Research Center, Baku, Azerbaijan
\item \Idef{org90}National Research Centre Kurchatov Institute, Moscow, Russia
\item \Idef{org91}Niels Bohr Institute, University of Copenhagen, Copenhagen, Denmark
\item \Idef{org92}Nikhef, Nationaal instituut voor subatomaire fysica, Amsterdam, Netherlands
\item \Idef{org93}Nuclear Physics Group, STFC Daresbury Laboratory, Daresbury, United Kingdom
\item \Idef{org94}Nuclear Physics Institute, Academy of Sciences of the Czech Republic, \v{R}e\v{z} u Prahy, Czech Republic
\item \Idef{org95}Oak Ridge National Laboratory, Oak Ridge, Tennessee, United States
\item \Idef{org96}Petersburg Nuclear Physics Institute, Gatchina, Russia
\item \Idef{org97}Physics Department, Creighton University, Omaha, Nebraska, United States
\item \Idef{org98}Physics department, Faculty of science, University of Zagreb, Zagreb, Croatia
\item \Idef{org99}Physics Department, Panjab University, Chandigarh, India
\item \Idef{org100}Physics Department, University of Cape Town, Cape Town, South Africa
\item \Idef{org101}Physics Department, University of Jammu, Jammu, India
\item \Idef{org102}Physics Department, University of Rajasthan, Jaipur, India
\item \Idef{org103}Physikalisches Institut, Eberhard Karls Universit\"{a}t T\"{u}bingen, T\"{u}bingen, Germany
\item \Idef{org104}Physikalisches Institut, Ruprecht-Karls-Universit\"{a}t Heidelberg, Heidelberg, Germany
\item \Idef{org105}Physik Department, Technische Universit\"{a}t M\"{u}nchen, Munich, Germany
\item \Idef{org106}Research Division and ExtreMe Matter Institute EMMI, GSI Helmholtzzentrum f\"ur Schwerionenforschung GmbH, Darmstadt, Germany
\item \Idef{org107}Rudjer Bo\v{s}kovi\'{c} Institute, Zagreb, Croatia
\item \Idef{org108}Russian Federal Nuclear Center (VNIIEF), Sarov, Russia
\item \Idef{org109}Saha Institute of Nuclear Physics, Kolkata, India
\item \Idef{org110}School of Physics and Astronomy, University of Birmingham, Birmingham, United Kingdom
\item \Idef{org111}Secci\'{o}n F\'{\i}sica, Departamento de Ciencias, Pontificia Universidad Cat\'{o}lica del Per\'{u}, Lima, Peru
\item \Idef{org112}SSC IHEP of NRC Kurchatov institute, Protvino, Russia
\item \Idef{org113}Stefan Meyer Institut f\"{u}r Subatomare Physik (SMI), Vienna, Austria
\item \Idef{org114}SUBATECH, IMT Atlantique, Universit\'{e} de Nantes, CNRS-IN2P3, Nantes, France
\item \Idef{org115}Suranaree University of Technology, Nakhon Ratchasima, Thailand
\item \Idef{org116}Technical University of Ko\v{s}ice, Ko\v{s}ice, Slovakia
\item \Idef{org117}Technical University of Split FESB, Split, Croatia
\item \Idef{org118}The Henryk Niewodniczanski Institute of Nuclear Physics, Polish Academy of Sciences, Cracow, Poland
\item \Idef{org119}The University of Texas at Austin, Physics Department, Austin, Texas, United States
\item \Idef{org120}Universidad Aut\'{o}noma de Sinaloa, Culiac\'{a}n, Mexico
\item \Idef{org121}Universidade de S\~{a}o Paulo (USP), S\~{a}o Paulo, Brazil
\item \Idef{org122}Universidade Estadual de Campinas (UNICAMP), Campinas, Brazil
\item \Idef{org123}Universidade Federal do ABC, Santo Andre, Brazil
\item \Idef{org124}University of Houston, Houston, Texas, United States
\item \Idef{org125}University of Jyv\"{a}skyl\"{a}, Jyv\"{a}skyl\"{a}, Finland
\item \Idef{org126}University of Liverpool, Liverpool, United Kingdom
\item \Idef{org127}University of Tennessee, Knoxville, Tennessee, United States
\item \Idef{org128}University of the Witwatersrand, Johannesburg, South Africa
\item \Idef{org129}University of Tokyo, Tokyo, Japan
\item \Idef{org130}University of Tsukuba, Tsukuba, Japan
\item \Idef{org131}Universit\'{e} Clermont Auvergne, CNRS/IN2P3, LPC, Clermont-Ferrand, France
\item \Idef{org132}Universit\'{e} de Lyon, Universit\'{e} Lyon 1, CNRS/IN2P3, IPN-Lyon, Villeurbanne, Lyon, France
\item \Idef{org133}Universit\'{e} de Strasbourg, CNRS, IPHC UMR 7178, F-67000 Strasbourg, France, Strasbourg, France
\item \Idef{org134}Universit\`{a} degli Studi di Pavia, Pavia, Italy
\item \Idef{org135}Universit\`{a} di Brescia, Brescia, Italy
\item \Idef{org136}V.~Fock Institute for Physics, St. Petersburg State University, St. Petersburg, Russia
\item \Idef{org137}Variable Energy Cyclotron Centre, Kolkata, India
\item \Idef{org138}Warsaw University of Technology, Warsaw, Poland
\item \Idef{org139}Wayne State University, Detroit, Michigan, United States
\item \Idef{org140}Wigner Research Centre for Physics, Hungarian Academy of Sciences, Budapest, Hungary
\item \Idef{org141}Yale University, New Haven, Connecticut, United States
\item \Idef{org142}Yonsei University, Seoul, Republic of Korea
\item \Idef{org143}Zentrum f\"{u}r Technologietransfer und Telekommunikation (ZTT), Fachhochschule Worms, Worms, Germany
\end{Authlist}
\endgroup
  %%%%%%% get the latest version before submitting
%
%\input{}              %%%%%%%%%%%% put the references here
%
\end{document}